\documentclass[pra,twocolumn,showpacs,aps]{revtex4-1}
\usepackage{color}
\usepackage{amsmath,amssymb}
\usepackage{soul}
\usepackage{txfonts}
\usepackage{graphicx}
\usepackage{subfigure}
\usepackage{placeins}
\usepackage[unicode]{hyperref}
\hypersetup{
    pdftitle={KZM winding},
    pdfauthor={R. G. McDonald and A. S. Bradley},
    pdfsubject={KZM winding},
    colorlinks=true,
    linkcolor=blue,
    citecolor=blue,
    filecolor=black,
    urlcolor=blue
}
\usepackage{bm}
\usepackage{epstopdf}
\usepackage{enumerate}
\def\urlprefix{}
   \def\url#1{}
\newcommand{\ve}{\varepsilon}

\newcommand{\rC}{\textbf{C}}
\newcommand{\rI}{\textbf{I}}

\newcommand{\EQ}[1]{\begin{align}#1\end{align}}

\newcommand{\eref}[1]{(\ref{#1})}
\newcommand{\fref}[1]{Fig.~\ref{#1}}

\begin{document}
\title{Reservoir interactions during Bose-Einstein condensation: \\
modified critical scaling in the Kibble-Zurek mechanism of defect formation}
\author{R. G. McDonald and A. S. Bradley} 
\affiliation{Department of Physics, QSO | Centre for Quantum Science, and Dodd-Walls Centre for Photonic and Quantum Technologies, University of Otago, Dunedin 9010, New Zealand.}
%\date{\today}
\begin{abstract}
As a test of the Kibble-Zurek mechanism (KZM) of defect formation, we simulate the Bose-Einstein condensation transition in a toroidally confined Bose gas using the stochastic projected Gross-Pitaevskii equation (SPGPE), with and without the energy-damping reservoir interaction. Energy-damping alters the scaling of the winding number distribution with the quench time - a departure from the universal KZM theory that relies on equilibrium critical exponents. Numerical values are obtained for the correlation-length critical exponent $\nu$ and the dynamical critical exponent $z$ for each variant of reservoir interaction theory. The energy-damping reservoir interactions cause significant modification of the dynamical critical exponent of the phase transition, whilst preserving the essential KZM critical scaling behavior. Comparison of numerical and analytical two-point correlation functions further illustrates the effect of energy damping on the correlation length during freeze out.
\end{abstract}
\maketitle

\section{Introduction}
The Kibble-Zurek mechanism (KZM) describes defect formation in the symmetry-breaking dynamics of a system undergoing a second-order phase transition \cite{Kibble:1976fm,Kibble:1980bw}. The theory exploits critical slowing down near the transition, whereby the system relaxation time diverges and the system becomes essentially frozen, allowing the description of critical \emph{dynamics} in terms of \emph{equilibrium} critical exponents~\cite{delCampo:2014eu}; thus the quench phenomena are set by the universality class of the system. The critical exponents determine the density of defects introduced by symmetry breaking during the quench, and thus are central to the testable predictions of KZM. Bose-Einstein condensation (BEC) is a $U(1)$ symmetry breaking transition whereby the phase of the order parameter acquires independent values over finite-sized domains, the size of which depends on the speed of the quench. Theoretical treatments of KZM in Bose-Einstein condensation have used a description of the Bose gas that can be reduced to a form of stochastic Ginzburg-Landau (G-L) theory~\cite{Laguna:1997jn,Yates:1998js,Antunes:1999cr,Bettencourt:2000ce,Weiler:2008eu,Damski10a,Das:2012ki,Su:2013dh}, involving a Gross-Pitaevskii equation coupled to a grand canonical reservoir providing a source of particles and thermal noise. Fundamentally, the model is a \emph{mean field theory} driven by the simplest possible reservoir coupling, and analyses of defect formation confirm the equilibrium scaling hypothesis of KZM. A basic question then arises: what is the role of a specific system's reservoir interactions in determining the critical dynamics during a quench?
\par
 A rigorous and tractable reservoir interaction theory has been developed for the dilute Bose gas from first principles in the form of the stochastic projected Gross-Pitaevskii equation (SPGPE)~\cite{Gardiner:2003bk,Bradley:2008gq,Rooney:2014kc,Rooney:2012gb,Bradley:2014a}. The theory is a synthesis of quantum kinetic theory~\cite{QKIII} and the projected Gross-Pitaevskii equation~\cite{Davis2001b}, and provides a tractable approach for numerical simulations of critical dynamics that includes all significant reservoir interaction processes. 
The SPGPE describes the evolution of a high-temperature partially condensed system within a classical field approximation, is valid on either side of the critical point, and in 3D has been used to quantitatively model the phase transition~\cite{Weiler:2008eu}, and high-temperature dynamics~\cite{Rooney:2013ff} observed in experiments. The complete SPGPE includes a number-damping reservoir interaction (G-L type) described in previous works~\cite{Weiler:2008eu,Rooney:2013ff}, and an additional interaction involving exchange of energy with the reservoir~\cite{Anglin:1999fn}; the latter is a number-conserving interaction that can have a significant influence on system evolution far from equilibrium~\cite{Blakie:2008is,Rooney:2012gb,Bradley:2015wb}, such as occurs in the region of critical slowing down near the transition.

In this work we investigate the effect of the energy damping reservoir interaction on the outcome of quenches across the Bose-Einstein condensation transition. We consider a toroidally trapped Bose gas consisting of a quasi-1D superfluid fraction immersed in a 3D thermal cloud. This system may be modelled using the effective 1D-SPGPE description for elongated systems~\cite{Bradley:2015wb}. A finite persistent current circulating around the ring provides a clear signature of symmetry breaking~\cite{Zurek:1985wh}. We perform quenches of the reservoir chemical potential for a range of quench times, with and without the energy damping terms, to produce a statistical distribution of the final winding number of the persistent current. The Kibble-Zurek mechanism predicts a power law relation between the quench time and the standard deviation of the final winding number. We numerically determine the power-law exponent for these relations using the two reservoir interaction theories and compare with the predictions of mean field theory. Inclusion of energy damping causes the power-law exponents to depart from the mean field values. In particular, the dynamical critical exponent extracted from the complete SPGPE simulations differs from that predicted by mean field theory, does not conform to a known universality class, and suggests non-universal modifications to the critical dynamics. The effects of energy damping are further exemplified by comparing the dynamics of condensate number and two-point correlations for the two theories.
\section{Theory}
\subsection{The Kibble-Zurek mechanism}
%--------------------------------------------------------------------------------------------------------
\begin{figure}[t!]{
\begin{center} 
\includegraphics[width=.8\columnwidth]{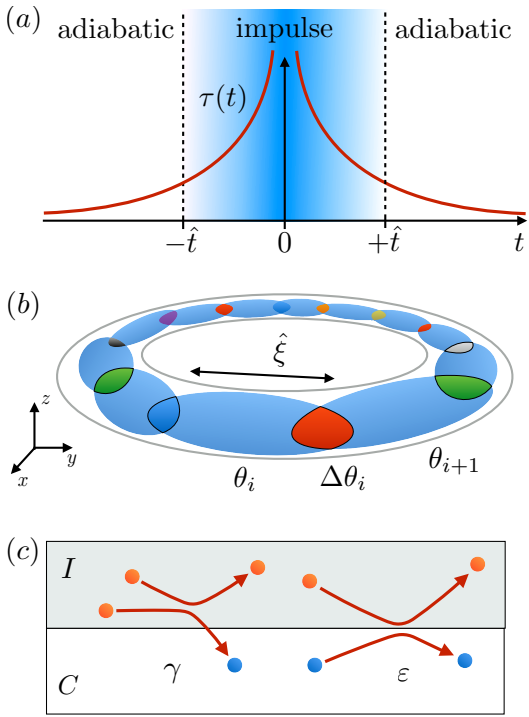}
\caption{(colour online) (a) Adiabatic-impulse model of dynamical symmetry breaking in KZM. The boundary between regions $\hat t$ is defined as the instant when the time remaining until the critical point is equal to the system relaxation time, i.e. when $\hat t=\tau(-\hat t)$. (b) Schematic of defect formation during Bose-Einstein condensation in a ring trap. The scale of regions acquiring a $U(1)$ symmetry-broken phase is set by the coherence length at the freeze-out time, $\hat\xi=\xi(-\hat t)$. Phase differences between adjacent domains, $\Delta \theta_i$, are resolved through the formation of defects. (c) Reservoir interactions in the stochastic projected Gross-Pitaevskii theory of the dilute Bose gas. The number-damping ($\gamma$) interaction drives condensate growth. The number-conserving energy damping process ($\varepsilon$) has a significant role far from equilibrium.
\label{Fig:schem}}
\end{center}}
\end{figure}
%--------------------------------------------------------------------------------------------------------
We first review aspects of KZM relevant for this work. We consider a system driven across a second order phase transition by a quench of the chemical potential from $-\mu_0$ to $\mu_0$. We define the \emph{reduced parameter} $\epsilon(t)$
\EQ{
\epsilon(t)&=\frac{\mu(t)}{\mu_0} = \frac{t}{\tau_Q}
\label{quench}
}
where $2\tau_Q$ is the quench duration and $t\in\left[-\tau_Q,\tau_Q\right]$. The equilibrium correlation length and dynamical relaxation time are related to the reduced parameter by
\EQ{
\xi(t)&=\frac{\xi_0}{|\epsilon(t)|^\nu},\label{EQ:cl}
}
and
\EQ{
\tau(t)&=\frac{\tau_0}{|\epsilon(t)|^{z\nu}},\label{EQ:rt}
}
respectively, where $\nu$ and $z$ are critical exponents defining the universality class of the phase transition, and $\xi_0$ and $\tau_0$ are constants that depend on the microscopic details of the system. Initially the system follows the quench adiabatically. However, as the system approaches the critical point, $\mu=0$,  the relaxation time diverges and there is thus an instant during the quench where the relaxation time is equal to the time remaining reach the critical point. This time $t=-\hat t$ is called the \emph{freeze-out time}, after which the system cannot keep up with the external parameter, and thus remains ``frozen" (impulse regime) until time $+\hat t$ after the transition when it can again respond to the environment, as shown in \fref{Fig:schem} (a). The freeze-out time satisfies the equation
\EQ{
\hat{t} &\equiv \tau(-\hat{t}),
} 
which may be solved \cite{Zurek:1985wh} to give
\EQ{
\hat{t}&=\left(\tau_0\tau_Q^{z\nu}\right)^{\frac{1}{1+z\nu}}=\tau_0^{1-\alpha} \tau_Q^\alpha.
\label{freeze}
}
where
\EQ{
\alpha&\equiv\frac{z\nu}{1+z\nu}.
\label{EQ:alpha}}
\par
The adiabatic-impulse approximation assumes that the system ceases to follow the equilibrium solution adiabatically precisely at $-\hat{t}$, when the dynamics are frozen until the system returns to adiabatic following at $\hat{t}$.
Using the freeze-out time we can obtain the freeze-out chemical potential and correlation length
\EQ{
\hat{\mu}&=\mu_0\left(\frac{\tau_0}{\tau_Q}\right)^{\frac{1}{1+z\nu}},
\label{chem}
}
and
\EQ{
\hat{\xi}&=\xi_0\left(\frac{\tau_Q}{\tau_0}\right)^{\frac{\nu}{1+z\nu}}.
\label{corr}
}
The density of topological defects can be estimated as the size of the symmetry broken domains at the freeze-out time [see \fref{Fig:schem} (b)], and is then given by
\EQ{
n&=\frac{\hat{\xi}^d}{\hat{\xi}^D}=\frac{1}{\xi_0^{D-d}}\left(\frac{\tau_0}{\tau_Q}\right)^{\frac{\nu}{1+z\nu}}
\label{ndef}
} 
where $D$ is the dimensionality of the system and $d$ is the dimensionality of the defects. This is typically an over-estimate of what is observed, and thus $\hat{\xi}$ is commonly replaced by $s\hat{\xi}$, where $s\sim\mathcal{O}(1-10)$ depends upon the model \cite{Laguna:1997jn,Yates:1998js,Antunes:1999cr,Das:2012ki}. From (\ref{freeze}) it is apparent that $\tau_Q$ may be chosen small enough such that the freeze-out time is larger than the total ramp time ($\hat{t}>\tau_Q$). This implies a fast quench limit, below which the system will effectively experience a jump in the chemical potential, rather than a ramp. To be in the regime where KZM scaling should apply, the quench time must satisfy $\tau_Q>\tau_0$.
\par
In a 1D Bose gas, the defects that form as a result of merging domains are grey solitons~\cite{Damski10a,Lamporesi:2013bi}. According to \eref{ndef}, the density of solitons after a quench should scale as
\EQ{
n_s&\propto\tau_Q^{\frac{-\nu}{1+z\nu}}.
}
The problem arises that grey solitons are unstable \cite{Burger:1999ew,Muryshev2002a,Jackson:2007gy,Damski10a}, and thus in the presence of any dissipation, the solitons will vanish following the quench. This inhibits verification of KZM via soliton counting as it is not clear how long after the quench to measure the number of solitons; it must be long enough that the solitons are formed such that they can be distinguished from density fluctuations due to noise, but short enough that the number of solitons has not appreciably decayed.
Fortunately, if the system has periodic boundary conditions, a net winding of the system can emerge post-quench as a topologically stable \cite{Kagan:2000gk} remnant of the domains that can also be used to test the theory \cite{Zurek:1985wh,Das:2012ki}
\par
We consider a 1D Bose gas in a toroid of circumference $L$. At the freeze-out time the domain size is given by (\ref{corr}), and thus the number of domains is $N\approx L/\hat{\xi}$. The probability distribution for the phase of one domain is uniform between $-\pi$ and $\pi$, with variance
\EQ{
\sigma^2(\theta)&=\int_{-\pi}^{\pi}\frac{\theta^2}{2\pi}d\theta=\frac{\pi^2}{3}.
}
The variance of the phase difference between two neighbouring domains $\Delta\theta_i=\theta_{i+1}-\theta_i$ is then $
\sigma^2(\Delta\theta_i)=2\pi^2/3$,
and the variance of the accumulated phase around the toroid is
\EQ{
\sigma^2(\theta_c)&=\sum_{N-1}\sigma^2(\Delta\theta_i)=(N-1)\frac{2\pi^2}{3}\approx N\frac{2\pi^2}{3} \approx \frac{L}{\hat{\xi}}\frac{2\pi^2}{3}.
}
The winding is related to the accumulated phase by $\mathcal{W}=\theta_c/2\pi$, and thus the variance of the winding is given by
\EQ{
\sigma^2(\mathcal{W})&=\frac{1}{(2\pi)^2}\sigma^2(\theta_c)\approx \frac{1}{6}\frac{L}{\hat{\xi}},
}
giving the scaling for the standard deviation of the winding distribution 
\EQ{
\sigma(\mathcal{W})&=\sqrt{\frac{L}{6\xi_0}}\tau_0^{\frac{\nu}{2(1+z\nu)}}\tau_Q^{\frac{-\nu}{2(1+z\nu)}}\propto \tau_Q^{-\beta},\label{EQ:std}
}
where
\EQ{
\beta&\equiv\frac{\nu}{2(1+z\nu)}.
\label{beta}
}
Following a quench, the system may be left to return to a stable state which may contain some non-zero winding. This winding can then be measured experimentally by interferometry \cite{Eckel:2014gf,Corman:2014cm}. The downside to measuring the winding over the defects themselves is that the variance of winding number is a higher order moment and thus more trajectories are required to obtain good statistics. Furthermore, the power-law exponent for the winding is half that for defects, making the scaling more susceptible to error. 
\subsection{SPGPE in one dimension}
The stochastic projected Gross-Pitaevskii equation (SPGPE) uses $\rC$-field methods to model Bose gases at finite temperature \cite{Gardiner:2003bk,Rooney:2012gb}. The modes of the system are divided into two distinct regions; the \emph{coherent region} (\rC) consisting of modes with energy less than a specified cutoff ($\epsilon_{\rm cut}$), and the \emph{incoherent region} (\rI) which contains the thermalised high-energy modes; the incoherent region acts as a thermal reservoir to the $\rC$-field. The SPGPE has been successfully numerically implemented \cite{Rooney:2012gb,Rooney:2014kc} despite its complexities. When considering systems that are tightly confined in one or more dimensions, the numerical implementation can be made more efficient by reducing the dimensionality of the SPGPE equations of motion~ \cite{Bradley:2015wb}.
\par
The 1D SPGPE is obtained by assuming the low-energy fraction of the system affording a SPGPE description is in the harmonic oscillator ground states for the two tightly trapped dimensions, enabling integrating over these dimensions \cite{Bradley:2015wb}. The transverse dimensions are confined by a parabolic trap with harmonic oscillator frequency $\omega_\perp$ and oscillator length $a_\perp=\sqrt{\hbar/m\omega_\perp}$. The resulting equation of motion is
\EQ{\label{1dspgpe}
(S)\hbar d\psi(x,t) ={}&{\cal P}_x\Bigg\{(i+\gamma)(\mu-\mathcal{L})\psi(x,t) dt+\hbar dW_\gamma(x,t)\nonumber\\
{}&-iV_\ve(x,t)\psi dt+i\hbar\psi(x,t) dW_\ve(x,t)\Bigg\}.
}
where $(S)$ denotes Stratonovich integration. The 1D projection operator $\mathcal{P}_x$ implements the energy cutoff $\epsilon_{\rm cut}$ in the remaining dimension $x$. The Hamiltonian evolution, with reservoir chemical potential as energy reference, is generated by
\EQ{\label{Ldef}
\mathcal{L}\psi(x,t)&\equiv\left[\mathcal{H}(x)+g_1|\psi(x,t)|^2-\mu \right]\psi(x,t)
}
where $\mathcal{H}(x)=-\hbar^2\partial_x^2/2m+V_{\rm ext}(x)$ is the single-particle Hamiltonian with external potential $V_{\rm ext}(x)$, and $g_1=2\hbar\omega_\perp a_s$ is the 1D interaction strength with s-wave scattering length $a_s$. The reservoir is described by chemical potential ($\mu$), temperature ($T$), and cutoff energy ($\epsilon_{\rm cut}$), and the functions
\begin{subequations}
\label{allsgpe}
\EQ{
\label{gamdef}
\gamma&=\frac{8a_s^2}{\lambda^2_{dB}}\sum_{j=1}^\infty\frac{e^{\beta\mu(j+1)}}{e^{2\beta\epsilon_{\rm cut}j}}\Phi\left[\frac{e^{\beta\mu}}{e^{2\beta\epsilon_{\rm cut}}},1,j \right],\\
\label{ve}
V_\ve(x,t)&=-\hbar\int dx' \ve(x-x')\partial_{x'}j(x',t),\\
\label{jdef}
j(x,t)&=\frac{i\hbar}{2m}\left[\psi\partial_x\psi^*-\psi^*\partial_x\psi\right],\\
\label{epsdef}
\ve(x)&=\frac{\mathcal{M}}{2\pi}\int dk\; e^{ikr} S_1(k),\\
\label{Sdef}
S_1(k)&\equiv {\rm erfcx}\left(\frac{|k|a_\perp}{\sqrt{2}}\right)(8\pi a_\perp^2)^{-1/2},\\
\label{Mdef}
\mathcal{M}&\equiv \frac{16\pi a_s^2}{e^{\beta(\epsilon_{\rm cut}-\mu)}-1},
}
\end{subequations}
where $\lambda_{dB} = \sqrt{2\pi\hbar^2/m k_BT}$, $\beta=1/k_BT$, $\Phi[z,x,a]=\sum_{k=0}^\infty z^k/(a+k)^x$ is the \emph{Lerch transcendent}, and 
\EQ{\label{erfcx}
{\rm erfcx}(x)&\equiv e^{x^2}{\rm erfc}(x)
}
is the \emph{scaled complementary error function}. The noise terms are Gaussian, with non-vanishing correlations
\EQ{
\label{nNoise}
\langle dW^*_\gamma(x,t)dW_\gamma(x',t)\rangle&=\frac{2\gamma k_BT}{\hbar}\delta(x,x')dt,\\
\label{eNoise}
\langle dW_\ve(x,t)dW_\ve(x',t)\rangle&=\frac{2k_BT}{\hbar}\ve(x-x')dt,
}
where $\delta(x,x')=\sum_{n\in\rC}\phi_n(z)\phi^*_n(x')$ is the $\delta$ function for the $\rC$ region.
\par
The first two terms on the RHS of (\ref{1dspgpe}) give the simple growth or \emph{number-damping} SPGPE defined by \eref{Ldef}, \eref{gamdef}, and describing particle exchange with the $\rI$-region [the PGPE is recovered as the special case $\gamma\equiv 0$]. The final two terms on the RHS of (\ref{1dspgpe}), are the \emph{energy damping} terms, describing energy-exchanging interactions between the $\rC$-field and the thermal cloud \emph{without particle exchange}. These terms involve a potential, \eref{ve}, and depend upon the divergence of the current \eref{jdef}, with strength determined by the rate function \eref{epsdef}, scattering kernel \eref{Sdef}, and amplitude \eref{Mdef}. Both processes involve an associated noise, \eref{nNoise}, \eref{eNoise}, and satisfy the fluctuation-dissipation theorem. The two reservoir interaction processes are shown schematically in \fref{Fig:schem} (c).
\subsection{Two-point correlations of the 1D SPGPE}
Although finding analytic solutions to the full 1D SPGPE (\ref{1dspgpe}) is very difficult, some analytic progress can be made for the number-damping SPGPE
\EQ{
\hbar d\psi(x,t)&={\cal P}_x\Bigg\{(i+\gamma)(\mu-\mathcal{L})\psi(x,t) dt+\hbar dW_\gamma(x,t)\Bigg\}.
}
We assume that we are on the symmetric side of the transition ($\mu<0$), allowing us to approximate $g_1|\psi|^2\approx0$, as the number density of the $\rC$-field is very low.
Transforming to momentum space we obtain the equation of motion
\begin{equation}
\hbar d\phi\left(k,t\right)=\left(i+\gamma\right)\left(\mu(t) -\frac{\hbar^2 k^2}{2m}\right)\phi\left(k,t\right) dt + \hbar dU\left(k,t\right)
\label{momeq}
\end{equation}
where
\EQ{
\phi\left(k,t\right)&=\left(2\pi\right)^{-1/2}\int dx e^{-ikx} \psi\left(x,t\right)
}
is the $\rC$-field wave function in momentum space and
\EQ{
dU\left(k,t\right)&=\left(2\pi\right)^{-1/2}\int dx e^{-ikx}dW_\gamma\left(x,t\right)
}
is the number damping noise in momentum space. The general solution to (\ref{momeq}) is
\EQ{
\phi\left(k,t\right)={}&e^{-\int_0^t ds \;\Lambda\left(k,s\right)}\phi\left(k,0\right) \nonumber\\
{}& + \int_0^t ds\; e^{-\int_{s}^t ds'\Lambda\left(k,s'\right)}dU\left(k,s\right),
\label{momsol}}
where
\EQ{
\Lambda(k,t)&=\left(i+\gamma\right)\frac{1}{\hbar}\left(\frac{\hbar^2k^2}{2m}+|\mu(t)|\right).
\label{lambda}}
We first consider the equilibrium system with a constant negative chemical potential.  The coefficient (\ref{lambda}) is then time-independent, making the time-integrals in (\ref{momsol}) trivial. Taking $\langle\phi^*\left(k,t\right)\phi\left(k',t'\right)\rangle $ from (\ref{momsol}) and letting $t,t' \rightarrow \infty$, while keeping $|t-t'|$ constant and finite, we then transform back to position space to obtain the stationary two-point correlation function
\EQ{\label{steadsol}
\langle\psi^*\left(x,t\right)\psi\left(x',t'\right)\rangle_s&=\frac{k_B T}{2|\mu|\xi} G\left(\frac{|x-x'|}{\xi},(1-i/\gamma)\frac{|t-t'|}{\tau}\right),
}
%--------------------------------------------------------------------------------------------------------
\begin{figure*}[t!]{
\begin{center}
 \includegraphics[trim = 10mm 0mm 0mm 0mm, width=0.8\paperwidth]{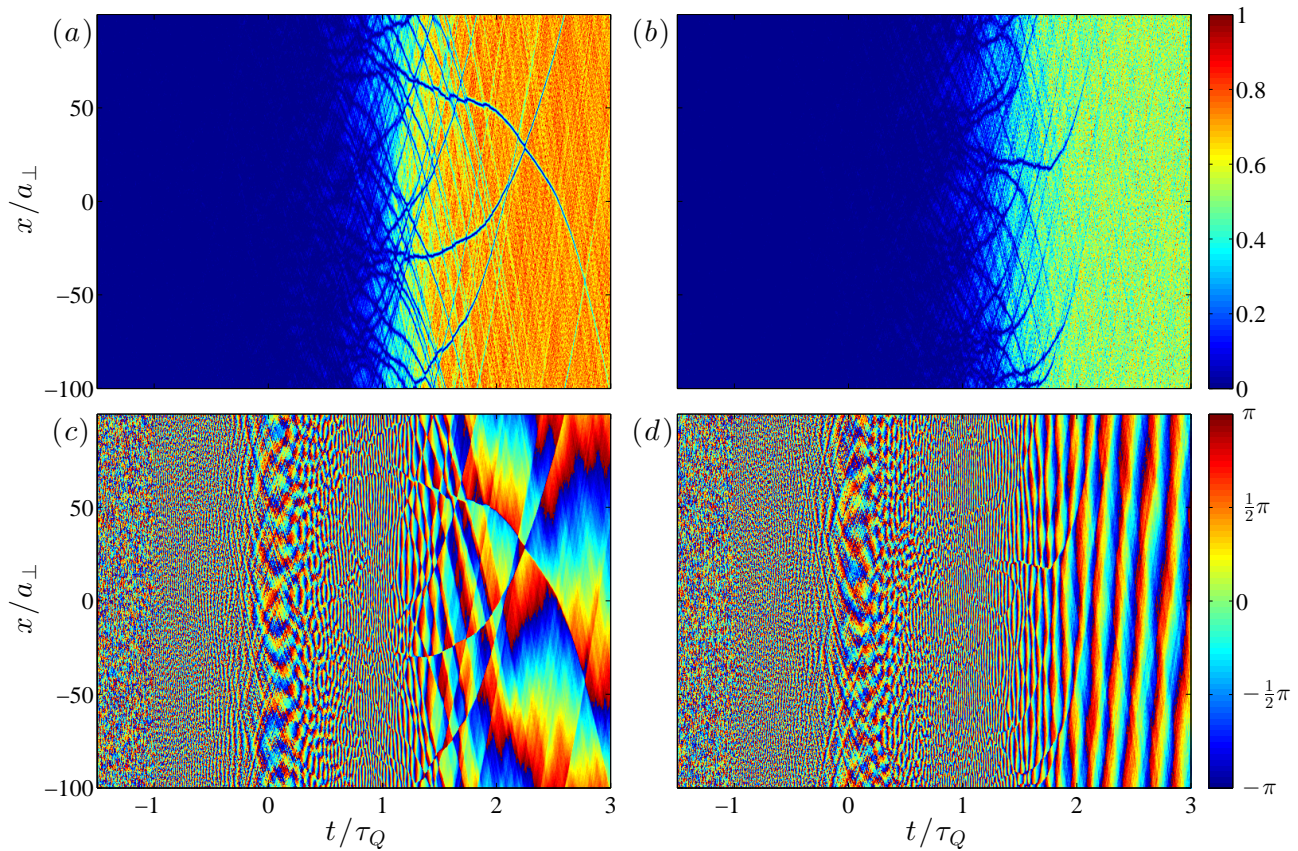}
\caption{(colour online) Single trajectories of the $\rC$-field during a quench ($\tau_Q=e^5 \omega_\perp^{-1}$), showing (a) number density and (c) phase for the number-damping SPGPE, and (b) number density and (d) phase for the full SPGPE. During initial density growth there are numerous solitons, evident as low-density notches with an associated phase jump, that quickly decay leaving a persistent current. The decay of solitons appears to be more rapid for the full SPGPE simulations.
\label{solitons}}
\end{center}}
\end{figure*}
%--------------------------------------------------------------------------------------------------------
where 
\EQ{
G(x,t)\equiv{}& \frac{1}{\pi}\int dk\;e^{-ikx}\;\frac{e^{-(1+k^2)t}}{1+k^2}\nonumber\\\label{Gdef}
={}&\frac{1}{2}e^{-2t}\left[e^{2t-x}{\rm erfc}\left(\frac{2t-x}{2\sqrt{t}}\right)+e^{2t+x}{\rm erfc}\left(\frac{2t+x}{2\sqrt{t}}\right)\right],\;\;\;\;\;
}
and we have identified the steady state correlation length
\EQ{\label{EQ:ecl}
\xi&\equiv\sqrt{\frac{\hbar^2}{2m|\mu|}},}
and relaxation time
\EQ{\label{EQ:ert}
\tau&\equiv\frac{\hbar}{\gamma|\mu|}.}
To verify that this expression has the correct limits, we see that $G(0,t)={\rm erfc}(\sqrt{t})$,
so that using ${\rm erfc}(z)\to e^{-z^2}/\sqrt{\pi}z$ for large $|z|$ gives
\EQ{\label{Gtlim}
\langle\psi^*\left(x,t\right)\psi\left(x,t'\right)\rangle_s&\to\frac{k_B T}{2|\mu|\xi}\frac{e^{-(\gamma-i)|\mu||t-t'|/\hbar}}{\sqrt{\pi(\gamma-i)|\mu||t-t'|/\hbar}},
}
confirming that $\tau$ is the relaxation time. Similarly, $G(x,0)=e^{-|x|}$, and hence at equal times we recover
\EQ{\label{equaltime}
\langle\psi^*\left(x,t\right)\psi\left(x',t\right)\rangle_s&=
\langle\psi^*(x,t)\psi(x^\prime,t)\rangle=\frac{k_BT}{2|\mu|\xi}e^{-|x-x^\prime|/\xi},
}
confirming $\xi$ as the correlation length.
Comparing (\ref{EQ:ecl}) and (\ref{EQ:ert}) with (\ref{EQ:cl}) and (\ref{EQ:rt}) respectively reveals the values of the critical exponents for the number-damping SPGPE theory, namely the correlation length critical exponent $\nu=1/2$ and the dynamical critical exponent $z=2$. We also obtain the constants of proportionally for the scaling relations
\EQ{\xi_0&=\frac{\hbar}{\sqrt{2 m \mu_0}},\label{EQ:xi0}\\
 \tau_0&=\frac{\hbar}{\gamma \mu_0}.\label{EQ:tau0}
 }
We now turn to the dynamics of the linear quench described by \eref{quench} for which one may find the equal time correlation function on the symmetric side of the transition within the same linear approximation used above~\cite{Damski10a}. The number-damped SPGPE has formal solution \eref{momsol}, for \eref{lambda} with $\mu(t)=\mu_0\epsilon(t)$. Carrying out the integrals in \eref{momsol}, and taking the limit $|\mu_0|\to\infty$, we find that the equal-time correlation function can be written as
\EQ{\label{eqT}
\langle\psi^*\left(x,t\right)\psi\left(x',t\right)\rangle_s&=\frac{k_BT}{2|\hat{\mu}|\hat{\xi}}\;f\left(\frac{|x-x' |}{\hat{\xi}},\frac{\hat{\xi}}{\xi(t)}\right),
}
where
\EQ{\label{fdef}
f(x,y)&=\frac{1}{\sqrt{\pi}}\int dk\;e^{ikx}{\rm erfcx}\left(k^2+y^2\right),
}
and $\hbar^2/2m\xi(t)^2\equiv |\mu(t)|$.
In the early part of the quench $\xi(t)\ll \hat\xi$, and we can use ${\rm erfc}(z)\to e^{-z^2}/\sqrt{\pi}z$ to find the asymptotic form for $y\gg 1$
\EQ{\label{bigy}
f(x,y)&\to\frac{1}{y^2 \pi }\int dk\;e^{ikx}\frac{1}{1+k^2/y^2}=\frac{e^{-|x|y}}{y},
}
and 
\EQ{\label{adR}
\langle\psi^*(x,t)\psi(x^\prime,t)\rangle&=\frac{k_BT}{2|\mu(t)|\xi(t)}e^{-|x-x^\prime|/\xi(t)},
}
the adiabatic form expected from~\eref{equaltime}. In the impulse regime \eref{fdef} is transcendental and must be numerically evaluated, however, the dependence upon only $|x-x'|/\hat\xi$ is a clear signature of universal scaling in the approach to the critical point. Numerically, one finds that in the freeze-out regime $f(x,y)$ depends only very weakly on $y$ in the neighbourhood $y\lesssim 1$, and the functional form freezes into an impulse regime for $0\leq y\leq1$. For $\mu>0$ ($1<y$) we must rely on numerical simulations of the SPGPE.
\section{Simulations}
\subsection{SPGPE simulations}
We consider a system with toroidal geometry, where the radial extent of the system is much smaller than the circumference of the toroid; the transverse trapping frequency is $\omega_\perp/2\pi=200$Hz, while the circumference is $L=200 a_\perp$ where $a_\perp$ is the transverse harmonic oscillator width. The dimensionally reduced system is then equivalent to a partially degenerate 1D Bose gas in a homogeneous trap of length $L$ with periodic boundary conditions, embedded in a 3D thermal cloud. We use atomic parameters for $^{87}{\rm Rb}$, giving the 1D interaction parameter $g_1 = 0.0139\hbar\omega_\perp a_\perp$.  We chose a grid of $M=1024$ points, which gives $\approx 4$ points per correlation length prior to and following the quench (the correlation length takes its smallest value prior to the quench). The chemical potential is quenched from $\mu = -\hbar\omega_\perp$ to $\mu = \hbar\omega_\perp$ i.e. $\mu_0 = \hbar\omega_\perp$. The temperature is held constant throughout all simulations at $T=0.5T_c$, where $T_c$ is the transition temperature for the ideal Bose gas confined to a 3D toroidal trap:
 %--------------------------------------------------------------------------------------------------------
\begin{figure}[t!]{
\begin{center}
 \includegraphics[trim = 0mm 0mm 0mm 0mm, width=\columnwidth]{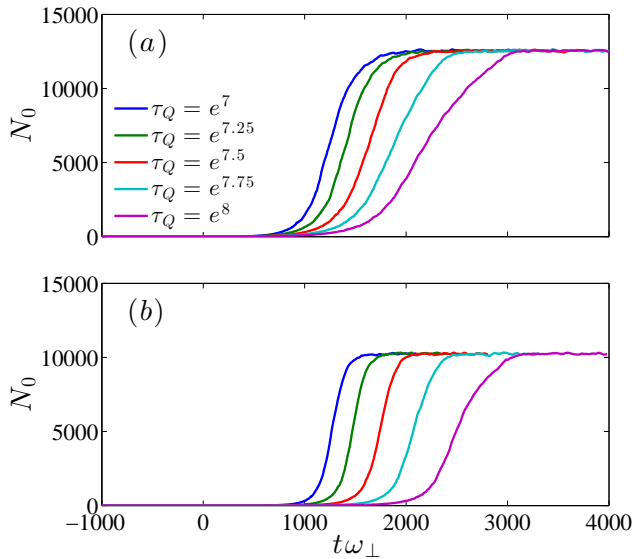}
\caption{(colour online) The condensate number dynamics for several quench times, using (a) the number-damping SPGPE, and (b) the full SPGPE.  The condensate number was extracted from the numerical data using the Penrose-Onsager criterion [see \eref{POcrit}].
\label{n0_unscaled}}
\end{center}}
\end{figure}
%--------------------------------------------------------------------------------------------------------
\EQ{T_c&=\frac{\hbar\bar{\omega}}{k_B}\left(\frac{N}{\zeta(5/2)}\right)^{2/5}
}
with $\bar{\omega}$ the geometric mean frequency of the toroid
\EQ{
\bar{\omega}^5&=\frac{2\pi\hbar}{mL^2}\omega_\perp^4,
}
$\zeta(z)$ the Riemann zeta function, and $N$ the number of particles~\cite{Bradley:2009bd}. To determine $T_c$ we have used the Thomas-Fermi value of the particle number $N=\mu L/g_1$ corresponding to the post-quench parameters, giving a value of $N\approx14400$. For the number-damping SPGPE these parameters give an average $\rC$-field population of $N_\rC \approx 14300$ and condensate number $N_0 \approx 12600$ in equilibrium post-quench, while the inclusion of energy-damping reduces these values to $N_C \approx 13100$ and $N_0 \approx 10200$.
\par
To obtain a thermalized initial state we use the $\rC$-field wave function $\psi(x)=0$, that is then evolved using the 1D SPGPE with a high number-damping rate (\ref{gamdef}) of $\gamma = 1$ for 1000 units of the relaxation time (\ref{EQ:ert}) to allow the system to come to equilibrium with the thermal cloud in its pre-quench state. For the quench we set the number-damping rate to a value more suitable for dynamics, $\gamma = 10^{-2}$, and begin the chemical potential ramp (\ref{quench}). Once the quench is finished ($t=\tau_Q$) we allow the system to evolve for a further 10 units of the relaxation time, which we have seen is sufficient for any post-quench dynamics to cease. For the number-damping SPGPE, the energy-damping is not included, so an energy-damping rate \eref{Mdef} of $\mathcal{M} = 0$ is used for the entirety of the simulation. For the full SPGPE, we also include energy-damping terms, with rate $\mathcal{M} = \gamma=10^{-2}$ for the entirety of the dynamics. We note that $\gamma\sim{\cal M}$ for experimentally relevant parameters~\cite{Rooney:2012gb}.
%--------------------------------------------------------------------------------------------------------
\begin{figure}[t!]{
\begin{center} 
\includegraphics[trim = 0mm 0mm 0mm 5mm, width=\columnwidth]{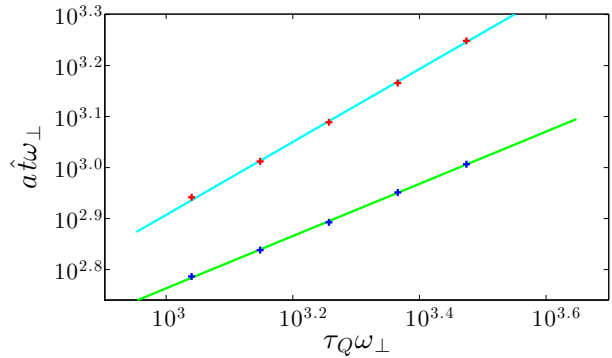}
\caption{(colour online) The results of our self-similarity algorithm with respect to the condensate number for the number-damping SPGPE and the full SPGPE. The numerical data for the number-damping (full) SPGPE is represented by blue (red) points, while the green (cyan) line is a least squares fit of a power law to the numerical data. The power law exponents are $\alpha=0.5119\pm0.0178$ and $\alpha=0.7145\pm0.0358$ for the number-damping SPGPE and full SPGPE respectively.
\label{Fig:SelfSim}}
\end{center}}
\end{figure}
%--------------------------------------------------------------------------------------------------------
\subsection{Results and analysis}
The qualitative behavior observed over the course of the quench is illustrated by the density and phase of the $\rC$-field $\psi(x,t)$, shown for example trajectories of the number-damping SPGPE and full SPGPE in \fref{solitons}; post-quench we see the emergence of decaying grey solitons in the density, and eventually a stable persistent current is evident in the phase. In order to analyse the SPGPE simulations further we first consider the condensate population, a quantity that can be computed throughout the quench dynamics in our 1D system, allowing extraction of the freeze-out time $\hat t$ numerically.
\par
We find the condensate mode and population using the Penrose-Onsager criterion \cite{Penrose:1956fr}, starting from the one-body density matrix 
\EQ{
\rho(x,x',t)&=\langle\psi(x,t) \psi^*(x',t) \rangle,
}
where the angled brackets denote ensemble averaging over trajectories; in the classical field regime this is equivalent to operator averaging as we may neglect commutators. Solving the eigenproblem
\EQ{\int dx' \rho(x,x',t) \phi_k(x',t) &= n_k(t) \phi_k(x,t).\label{POcrit}}
then gives the system orbitals $\phi_k(x)$ and their occupations $n_k$, with the largest eigenvalue giving the condensate number $N_0 \equiv \sup_k n_k$ and associated wave function $\phi_0(x)$. To construct the density matrix we performed $10^3$ trajectories per quench time using both the number-damping 1D SPGPE and full 1D SPGPE.
The mean condensate number for several quenches is shown in \fref{n0_unscaled}. The curves appear to be of the same functional form, but with rescaled time axis. This property is known as self-similarity, where rescaling the time axis by a particular factor will cause the curves to collapse onto a single curve; in this case the rescaling factor is the freeze-out time $\hat{t}$. 
\par
%--------------------------------------------------------------------------------------------------------
\begin{figure}[t!]{
\begin{center}
 \includegraphics[trim = 0mm 0mm 0mm 9mm, width=\columnwidth]{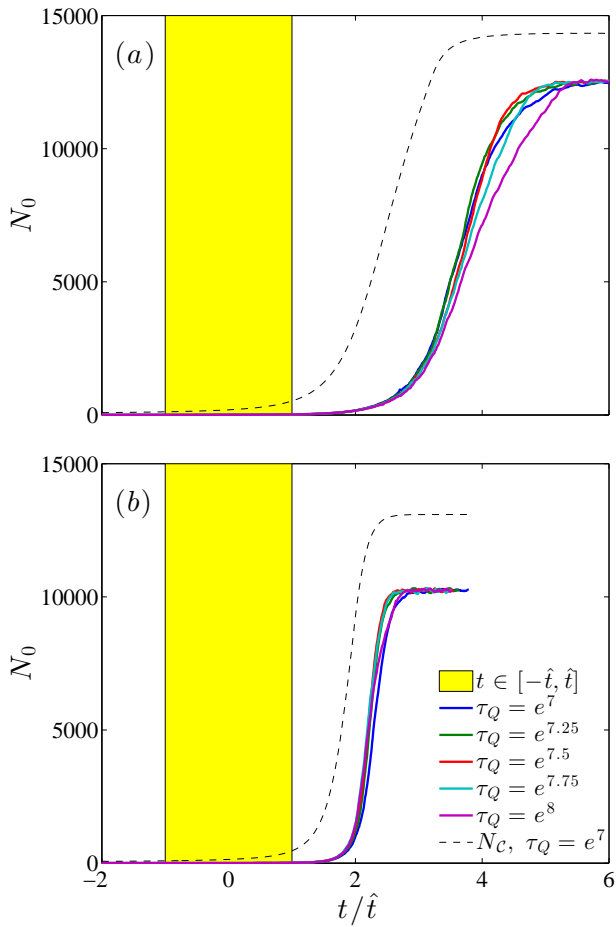}
\caption{(colour online) The condensate number over time for several quench times using (a) the number-damping SPGPE and, (b) the full SPGPE. The time axis has been scaled by the freeze-out time as predicted by our self-similarity algorithm for each theory. The yellow shaded region indicates the impulse regime where KZM approximates the system dynamics as frozen. For comparison, the total $\rC$-field population is also shown for one of the quenches (black dashed line).
\label{n0_scaled}}
\end{center}}
\end{figure}
%--------------------------------------------------------------------------------------------------------

%--------------------------------------------------------------------------------------------------------
\begin{figure}[t!]{
\begin{center}
 \includegraphics[trim = 0mm 0mm 0mm 5mm, width=0.9\columnwidth]{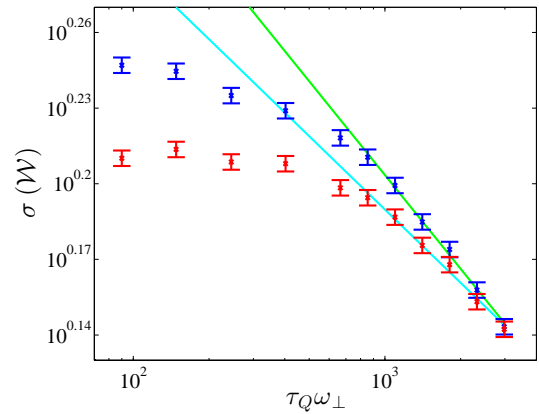}
\caption{(colour online) Standard deviation of the final winding for various quench times. The blue and red data points are the result of simulations of the number-damping SPGPE and the full SPGPE respectively. The green line is a a best fit for the number-damping SPGPE data in the regime that obeys a power law (\ref{EQ:std}), giving an exponent of $\beta=0.1236\pm0.0098$. The cyan line is the equivalent for the full SPGPE data, giving an exponent of $\beta=0.0966\pm0.0128$. 
\label{Fig:wind_stats}}
\end{center}}
\end{figure}
%--------------------------------------------------------------------------------------------------------
%--------------------------------------------------------------------------------------------------------
\begin{figure*}[t!]{
\begin{center}
 \includegraphics[trim = 0mm 0mm 0mm 0mm, width=0.75\paperwidth]{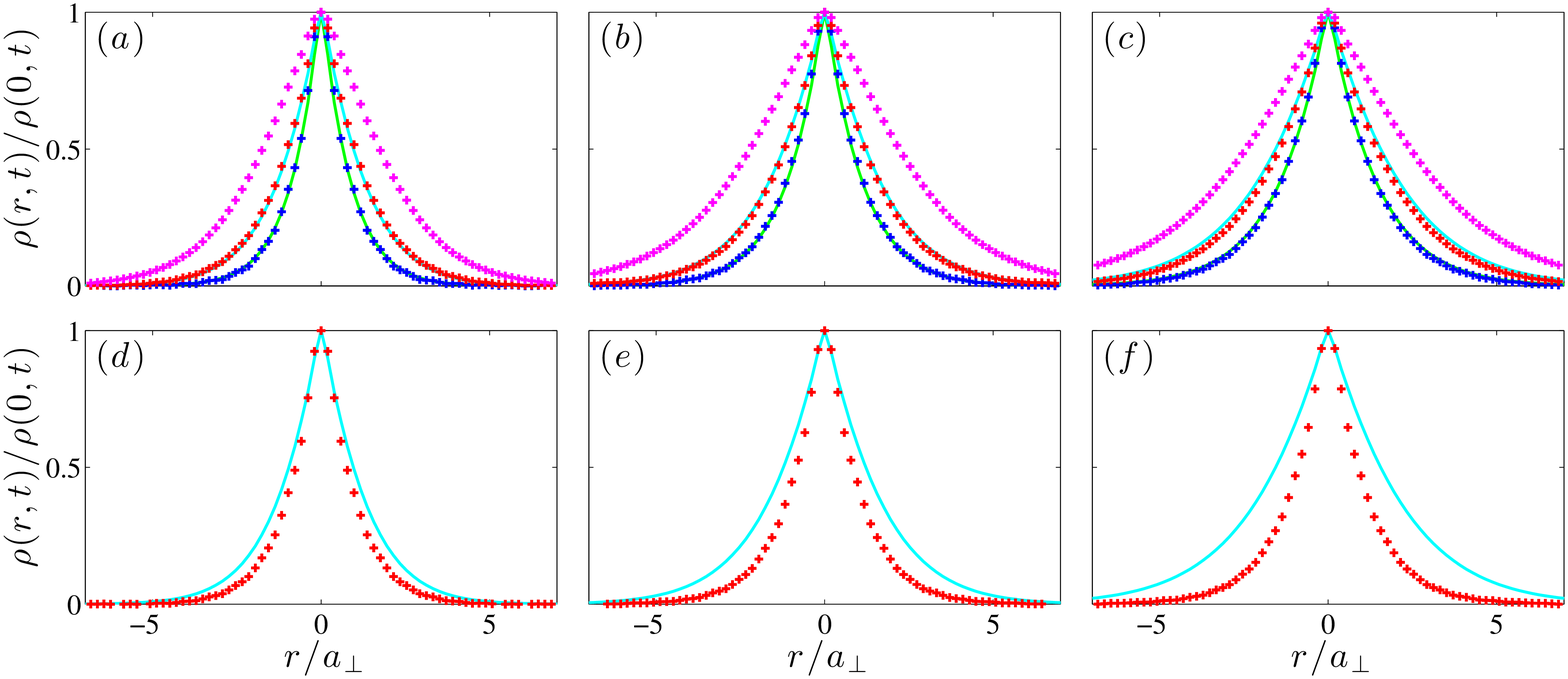}
\caption{(colour online) The two-point correlation function $\rho(r,t)=\langle\psi^*\left(x,t\right)\psi\left(x',t\right)\rangle$, where $r\equiv x-x'$, at various times for three different quench times simulated using both the number-damping SPGPE and the full SPGPE. The quench times are $\tau_Q \omega_\perp=e^6$ (a, d), $\tau_Q\omega_\perp=e^7 $ (b, e), and $\tau_Q \omega_\perp=e^8$ (c, f). The top row (a-c) shows the normalised numeric correlation function at $t=-\hat{t}$ (blue), $t=0$ (red), and $t=\hat{t}$ (magenta) resulting from simulations of the number-damping SPGPE, as well as the normalised analytic correlation function (\ref{eqT}) at $t=-\hat{t}$ (green) and $t=0$ (cyan). The bottom row (d-f) shows the normalised numeric correlation function at $t=0$ (red) resulting from simulations of the full SPGPE, as well as the normalised correlation function (\ref{eqT}) at $t=0$ (cyan).
\label{Fig:correlations}}
\end{center}}
\end{figure*}

%------------------------------------------------------------------------------------------------------------
\par
We find the freeze-out time from these curves based on their self-similarity. A reference condensate number is chosen relatively close to where the condensate number initially grows. The time at which this reference is reached should be linearly proportional to the freeze-out time, and thus obey a power law
\EQ{\label{alg}
a\hat{t}&=a\left(\tau_0 \tau_Q^{z\nu} \right)^{\frac{1}{1+z\nu}},
}
where $a$ is a constant of order unity. We then use a power-law fit to obtain the exponent $\alpha$ of (\ref{EQ:alpha}) and thus the value of $z\nu$. This allows us to determine the actual freeze-out time using (\ref{freeze}) with the constants $\xi_0$ (\ref{EQ:xi0}) and $\tau_0$ (\ref{EQ:tau0}).
\par
We use a reference condensate number of $N_0 = 100$, occurring rapidly after initial condensate growth. Figure \ref{Fig:SelfSim} shows the time at which the reference condensate number is reached against the quench time for several different quench times. Both theories give results that obey a power law, with exponent (\ref{EQ:alpha}) differing between the theories. A least-squares-fit to the results of simulations using the number-damping SPGPE gives $\alpha = 0.5119 \pm 0.0178$, consistent with the mean-field prediction $\alpha = 0.5$. A least-squares-fit for the full SPGPE gives $\alpha = 0.7145 \pm 0.0358$, a value that is inconsistent with mean field theory. 
\par
In Figure \ref{n0_scaled} we show the condensate number over time for the same quenches as Figure \ref{n0_unscaled}, this time with the time axis rescaled by the freeze-out time (\ref{freeze}) found using the values of $z\nu$ found above and the constants $\xi_0$ (\ref{EQ:xi0}) and $\tau_0$ (\ref{EQ:tau0}). The impulse regime $t\in[-\hat{t},\hat{t}]$ is highlighted to emphasise that condensate growth does not begin until the system has entered the adiabatic regime. Rescaling the time axis by $\hat{t}$ results in the curves lying on top of one another for both theories. Note that the value for $\tau_0$ used in \eref{alg}, given by Eq.~\eref{EQ:tau0}, is essentially exact for the number-damping SPGPE, while for the full SPGPE it is only an approximation as the additional damping terms change the relaxation time, however the power law extracted from Fig.~\ref{Fig:SelfSim} is unaltered by this choice. We have also included the growth of the $\rC$-field particle number $N_\rC(t)$ for one value of the quench time for comparison. 
\par
The final winding number is calculated using $\mathcal{W}=\theta_c/2\pi$ where $\theta_c$ is the accumulated phase of the final wave function around the toroid. For the purposes of obtaining the statistical distribution of the final winding number we performed $10^4$ trajectories per quench time using both the number-damping 1D SPGPE and the full SPGPE.
Figure \ref{Fig:wind_stats} shows the standard deviation of the final winding for a 
range of quench times. Far from the fast quench limit ($\tau_Q=10^2$) the winding standard deviation from simulations of both theories obeys a power law with respect to the quench time. The exponent (\ref{beta}) differs between the two; the number-damping SPGPE gives $\beta=0.1236\pm0.0098$, while the full SPGPE 
 gives $\beta=0.0966\pm0.0128$. The mean field value $\beta=0.125$ is within error of the number-damping SPGPE value, but not within error of the full SPGPE.
\par
The equations (\ref{EQ:alpha}) and (\ref{beta}) are a pair of simultaneous equations that relate $\alpha$ and $\beta$ to the critical exponents $\nu$ and $z$, giving
\EQ{z&=\frac{\alpha}{2\beta},\quad\nu=\frac{2\beta}{1-\alpha},}
and we can thus determine the values of $\nu$ and $z$ from our numerical simulations of quenches. The number-damping SPGPE gives the critical exponents $\nu=0.5065\pm0.0586$ and $z=2.071\pm0.236$. These are consistent with the equilibrium mean field critical exponents $\nu=1/2$ and $z=2$, a result that was also found in Ref.~\cite{Das:2012ki}. The full SPGPE give the critical exponents $\nu=0.6767\pm0.1745$ and $z=3.698\pm0.675$, a significant departure from the equilibrium mean field theory.
\par
We have also numerically calculated the two-point correlation function using ensembles of trajectories for several quench times; this is shown in Figure \ref{Fig:correlations}. For the number-damping SPGPE, we show the numerical correlation function at the times $t=\left\{-\hat{t},0,\hat{t}\right\}$ with the analytical correlation function (\ref{eqT}) at the times $t=\left\{-\hat{t},0\right\}$, where $\hat{t}$ is calculated using the constants $\xi_0$ (\ref{EQ:xi0}) and $\tau_0$ (\ref{EQ:tau0}) and the numerically obtained critical exponents $\nu$ and $z$. For the full SPGPE, we do not have values for the constants $\xi_0$ and $\tau_0$, and thus we cannot calculate the freeze-out time $\hat{t}$. Hence we only show the numerical correlation function at $t=0$ and compare this to the analytical form for the correlation function (\ref{eqT}) at the critical point ($t=0$). We see that for number-damping at the time $t=-\hat{t}$ and $t=0$, the numerical data shows excellent agreement with the analytical expression (\ref{eqT}), while at $t=\hat{t}$ the correlation length has clearly grown beyond what is predicted by (\ref{eqT}); this is unsurprising as (\ref{eqT}) was derived under the assumption that the transition has not yet been reached. For the full SPGPE we see that the correlation length is less than that predicted by (\ref{eqT}), possibly a consequence of the extra noise.
\section{Discussion}
Our results indicate that the energy-damping reservoir interaction can have a significant effect on the Bose-Einstein condensation transition. While the mean field critical exponents $\nu=1/2$ and $z=2$ were consistent with the number-damping SPGPE (as can also be shown analytically), the inclusion of the energy-damping terms resulted in an increase in both $\nu$ and $z$ to the point where they were no longer consistent with mean field theory. Other possible universality classes include the F-model \cite{Hohenberg:1977fq} ($\nu=2/3,\; z=3/2$), for which there has been experimental evidence \cite{Navon:2015jd}, or the 3D XY model ($\nu=0.6717,\;z=1.9550$), which is thought to be the class to which the BEC transition belongs \cite{Campostrini:2006io}. The value of the correlation-length critical exponent $\nu$ from our full SPGPE results is consistent with both these universality classes, however the value of the dynamical critical exponent $z$ is not. 
\par
Experimental temperature quenches have been performed in toroidally trapped BECs \cite{Corman:2014cm} finding power-law exponents in agreement with mean field theory. However, a recent experiment extracting the freeze-out correlation length \eref{corr} of a quenched  Bose gas in a box trap~\cite{Navon:2015jd} produced power-law exponents more consistent with the F-model \cite{Hohenberg:1977fq}. Our results suggest that the microscopic reservoir interactions, in particular the energy damping process, induce \emph{non-universal modifications} to the critical evolution. The modification is most evident in the dynamical critical exponent. Further work is needed to model specific experiments, and it may pose an experimental challenge to measure signatures of non-universal critical dynamics. In Table~\ref{Tab:exps} we summarize the power-law exponents for our simulations and relevant universality classes. 
\par
A further question remains as to the role of the values of the two damping rates $\gamma$ and $\mathcal{M}$; in this work we have assumed they time independent and equal. The precise ratio relevant for experiments, in principle also changing throughout the transition, may influence the final prediction for the critical exponents. A possible extension to this work would be to investigate the effects on the critical exponents when varying the ratio $\gamma/\mathcal{M}$; the true value of this ratio varies depending on the reservoir parameters of the experimental system of interest ($T$, $\mu$, $\epsilon_{\rm cut}$) \cite{Rooney:2012gb}.
\par
The energy damping terms affecting the condensate population may seem surprising given that the energy-damping process is number-conserving, however a previous investigation has shown that the energy-damping terms provoke a faster, more coherent approach to equilibrium \cite{Rooney:2012gb}, as is also consistent with the role of so-called \emph{scattering} terms in quantum kinetic theory~\cite{QKVI}. This more efficient equilibration may explain the lower final winding standard deviation upon inclusion of these terms, as the defects resulting from the phase transition are more efficiently damped away.
\par
%--------------------------------------------------------------------------------------------------------
\begin{table}[!t]
\begin{center}
    \begin{tabular}{ l|ccc}
    \hline
    Exponent  & $\alpha=z\nu/(1+z\nu)$ & $\beta=\nu/2(1+z\nu)$\\ \hline \hline
    Number-Damping SPGPE  & $0.5119\pm0.0178$ & $0.1236\pm0.0098$ \\ \hline
    Full SPGPE &  $\ 0.7145\pm0.0358$ & $0.0966\pm0.0128$\\ \hline
    Mean Field &  $0.5$& $0.125$ \\ \hline
    F-Model \cite{Hohenberg:1977fq} &  $ 0.5$ & $0.1667$ \\ \hline
    3D XY \cite{Campostrini:2006io} &  $0.5677$ & $0.1452$ \\ \hline \hline
    \end{tabular}
      \caption{The power law exponents for the freeze-out time (\ref{freeze}) and winding standard deviation (\ref{EQ:std}), for both the number-damping SPGPE and the full SPGPE. We have also included the values of these exponents as predicted by related universality classes. }
         \label{Tab:exps}
\end{center}
\end{table}
%--------------------------------------------------------------------------------------------------------
The presence of an additional noise source reduces the correlation length at the boundary of the impulse regime ($t=-\hat{t}$), reducing the domain size and increasing the number of defects. It would hence be informative to investigate the number of solitons over the course of the quenches, as is done in \cite{Damski10a}, and compare the two theories with experimental data~\cite{Lamporesi:2013bi}. This is a non-trivial task, as distinguishing solitons from density fluctuations in a noisy system can be difficult, particularly soon after the transition when the density is low. Long-time evolution to form a stable winding number suggests a coarsening dynamics process~\cite{Damle:1996ua,Bray:2002cb,Biroli:2010gna}.
\par
To the best of our knowledge we have given the most complete treatment of the reservoir interactions in the BEC transition, for the special case of an effectively 1D superfluid forming in a ring trap. Despite its microscopic foundation, certain details of experimental realizations are missing from our model. Most notably, our model of the quench ramps the $\rI$-region chemical potential, but includes no further dynamics of the $\rI$-region (thus far absent from any SPGPE theory), and we do not include a rigorous matching of the system to a specific set of experimental particle number measurements (as has been achieved in systems closer to equilibrium without fitting~\cite{Rooney:2013ff}, or in a quench by fitting the condensate growth rate to experimental data~\cite{Weiler:2008eu}), both of these aspects of modelling experimental quench dynamics remain open problems within SPGPE theory. 
\par
As this field is attracting increasing interest, it is of some value to connect the SPGPE model to related recent work.
We note that non-Markovian additive noise was shown to generate modified dynamical exponents in a G-L model~\cite{Bonart:2012cj}. 
Interestingly, quantum quenches may leave important signatures behind in higher order operator moments~\cite{Smacchia:2015cq}; such information is almost certainly absent from the classical field theory used in this work. As the BEC transition is dominated by classical fluctuations~\cite{Davis:2006ic}, the SPGPE nevertheless provides a quantitative theory of the transition.

The comparison of microscopic theories of dissipation with generic models also requires some comment. Recent work using the holographic duality approach~\cite{Sonner:2015dz} recovered mean-field power-laws for the winding number scaling in a model of a quenched superconducting transition. While the holographic method provides a very general approach with certain computational advantages, in holographic models the noise is introduced by hand via the fluctuation-dissipation theorem (FDT), generating dissipative terms and additive noise associated with particle growth. Numerical work typically focuses on a regime that is equivalent to overdamped BEC dynamics~\cite{Adams:2014be}, as discussed in Ref.~\cite{Billam:2015fj}. In contrast, the SPGPE theory is derived from a first-principles analysis of the Bose gas field theory, includes both additive and multiplicative noise (number damping and energy damping), and is underdamped; consistency with the FDT is an inherent property of the SPGPE. 

\section{Conclusions}
We have simulated chemical potential quenches across the Bose-Einstein condensation transition in a ring geometry by numerically solving the dimensionally reduced stochastic projected Gross-Pitaevskii equation with and without the energy damping terms. The final winding statistics of the number-damping SPGPE were found to obey the mean-field power-law exponent predictions of the Kibble-Zurek mechanism. The complete SPGPE, including energy damping terms, exhibits a modified power-law for the winding statistics. The freeze-out time was also extracted from the condensate number growth curves, with both theories again resulting in a power law with respect to the quench time but with differing exponents. While the number-damping SPGPE gave results consistent with mean field theory, we were unable to find a universality class with critical exponents consistent with the complete SPGPE results. The full SPGPE results are strongly suggestive of non-universal modifications to critical dynamics of the BEC phase transition, as is most clearly demonstrated by the modified dynamical exponent $z$. Our results highlight the importance of system-specific reservoir interactions in dynamical critical phenomena, as further suggested by recent experimental studies of the BEC phase transition~\cite{Lamporesi:2013bi,Corman:2014cm,Navon:2015jd}.
\acknowledgements
ASB is supported by a Rutherford Discovery Fellowship administered by the Royal Society of New Zealand.
%----------------------------------------------------------------------------------------------------------------------
%\section*{References}
%\bibliographystyle{apsrev4-1}
%\bibliography{PapersRefsComplete}

%Merlin.mbs v4.21 2009-07-09.
%

%-----------------------------------------------------------------------------------------------------------------------------------------------
\end{document}